\documentclass[11pt]{article}
\usepackage{moriond,epsfig}

\def\Journal#1#2#3#4{{#1} {\bf #2}, #3 (#4)}

\def\be{\begin{equation}}
\def\ee{\end{equation}}
\def\bea{\begin{eqnarray}}
\def\eea{\end{eqnarray}}

\newcommand{\f}{f^{(\nu)}}

\begin{document}
\vspace*{4cm} \title{DYNAMICAL MODELS AND NUMERICAL SIMULATIONS OF
INCOMPLETE VIOLENT RELAXATION}

\author{M.~TRENTI$^1$~\&~G.~BERTIN$^2$}

\address{(1)~Scuola Normale Superiore, piazza dei Cavalieri 7, I-56126 Pisa, Italy; \\(2)~Dipartimento di Fisica, Universit\`{a} di Milano, via
Celoria 16, I-20133 Milano, Italy}

\maketitle

\abstracts{N-body simulations of collisionless collapse have offered
important clues to the construction of realistic stellar dynamical
models of elliptical galaxies. Such simulations confirm and quantify
the qualitative expectation that rapid collapse of a self-gravitating
collisionless system, initially cool and significantly far from
equilibrium, leads to incomplete relaxation. In this paper we revisit
the problem, by comparing the detailed properties of a family of
distribution functions derived from statistical mechanics arguments
to those of the products of collisionless collapse found in N-body
simulations.}

\section{Introduction}
     
The collapse of a dynamically cold cloud of stars can lead to the
formation of realistic stellar systems, with projected density
profiles well represented by the $R^{1/4}$ law~\cite{van82}. The
theoretical framework for the mechanism of incomplete violent
relaxation that governs this scenario of structure formation was
proposed by Lynden-Bell~\cite{lyn67}, who argued that fast fluctuations
of the potential during collapse would lead to the formation of a
well-relaxed isotropic core, embedded in a radially anisotropic,
partially relaxed halo. This general picture served as a physical
justification for the construction of the so-called
$f_{\infty}$ models~\cite{ber84}, which indeed recovered the $R^{1/4}$ law and,
suitably extended to the case of two-component systems (to account for
the coexistence of luminous and dark matter), led to a number of
interesting applications to the observations~\cite{ber93}.

The application of statistical mechanics to this formation scenario~\cite{sti87}
also led to the derivation of a separate family of distribution
functions, the $\f$ models, which was recently shown to possess
interesting thermodynamic properties in the context of the gravothermal
catastrophe~\cite{ber03}. The key ingredient for the construction of the
$\f$ models is the conjecture that a \emph{third} quantity $Q$
(defined as $ Q = \int J^{\nu} |E|^{-3 \nu/4} f d^3q d^3p$), in
addition to the total mass $M$ and the total energy $E_{tot}$, is {\it
approximately} conserved during the process of collisionless
collapse. This quantity is introduced to model the process of {\it
incomplete} violent relaxation, ensuring a radially biased pressure
tensor and a $1/r^4$ density profile in the outer parts of the
system. A preliminary inspection of the general characteristics of the
$\f$ models then convinced us that, with significant advantage over
the $f_{\infty}$ models, they might also serve as a good framework to
interpret the results of simulations of collisionless collapse not
only qualitatively, but also in quantitative detail, which is
the main point of this paper.

\section{Models} \label{sec:fv}

The extremization of the Boltzmann entropy under the three constraints
described in the previous section leads to the distribution function
$f^{(\nu)} = A \exp {[- a E - d (J^2/|E|^{3/2})^{\nu/2}]}$, where $a$,
$A$, $d$, and $\nu$ are positive real constants; here $E$ ($E<0$) and $J$
denote single-star specific energy and angular momentum. At
fixed value of $\nu$, one may think of these constants as providing
two dimensional scales (for example, $M$ and $Q$) and one
dimensionless parameter, such as $\gamma = ad^{2/\nu}/(4 \pi GA)$.  In
the following we will focus on values of $\nu$ ranging from 3/8 to
1. The corresponding models are constructed by solving the Poisson
equation for the self-consistent mean potential $\Phi(r)$ generated by
the density distribution associated with $f^{(\nu)}$.  At fixed value
of $\nu$, the models thus generated make a one-parameter family of
equilibria, described by the concentration parameter $\Psi = -a
\Phi(0)$, the dimensionless depth of the central potential well. The
projected density profile of concentrated models is well fitted by the
$R^{1/n}$ law, with $n$ close to $4$, and the models provide
reasonable fits to the surface brightness and to the kinematic
profiles of bright elliptical galaxies~\cite{tre04}.

\section{Simulations of collisionless collapse}  \label{sec:init}
 
In principle, to study the process of collisionless collapse we have
two options: either a tree code~\cite{bar86} or a
particle-mesh~\cite{van82} algorithm. We are interested in the large
scale structure of the end-products of collisionless collapse, for
systems that do not exhibit large deviations from spherical
symmetry. The natural choice thus appears to be that of a
particle-mesh code, based on a spherical grid and an expansion in
spherical harmonics. The code used in the present study is thus a new
version~\cite{tren04} of the van Albada code. For completeness, we have
also run a number of comparison simulations with the fast code
developed by Dehnen~\cite{deh00}, which confirmed that our results do
not depend on the numerical scheme employed.

During the process of collisionless collapse, violent relaxation is
expected to wipe out much of the details that characterize the initial
conditions, so that the end-products are expected to have properties
that depend only or mostly on the initial value of the virial ratio $u
= (2K/|W|)_{t = 0}$, which sets the relevant collapse factor. In
reality, violent relaxation is incomplete. Therefore, the final state
is that of an approximate dynamical equilibrium characterized by an
anisotropic distribution function, different from a Maxwellian (which
would correspond to thermodynamic equilibrium). Because of such
incomplete relaxation, the end-products of the simulations do conserve
some memory of the initial state.

\begin{figure}
  \begin{center}
  \resizebox{130pt}{!}{\includegraphics{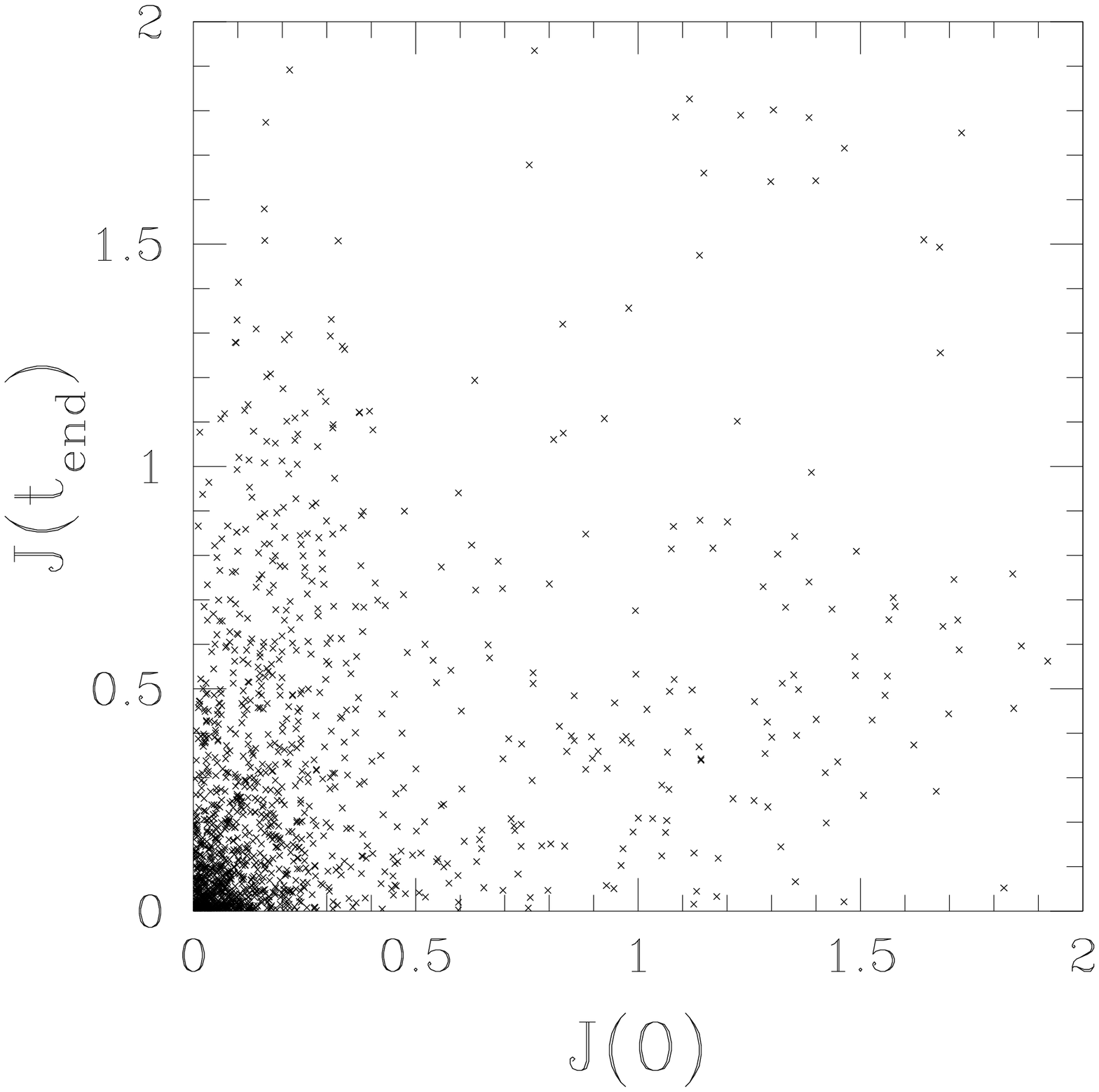}}
  \resizebox{130pt}{!}{\includegraphics{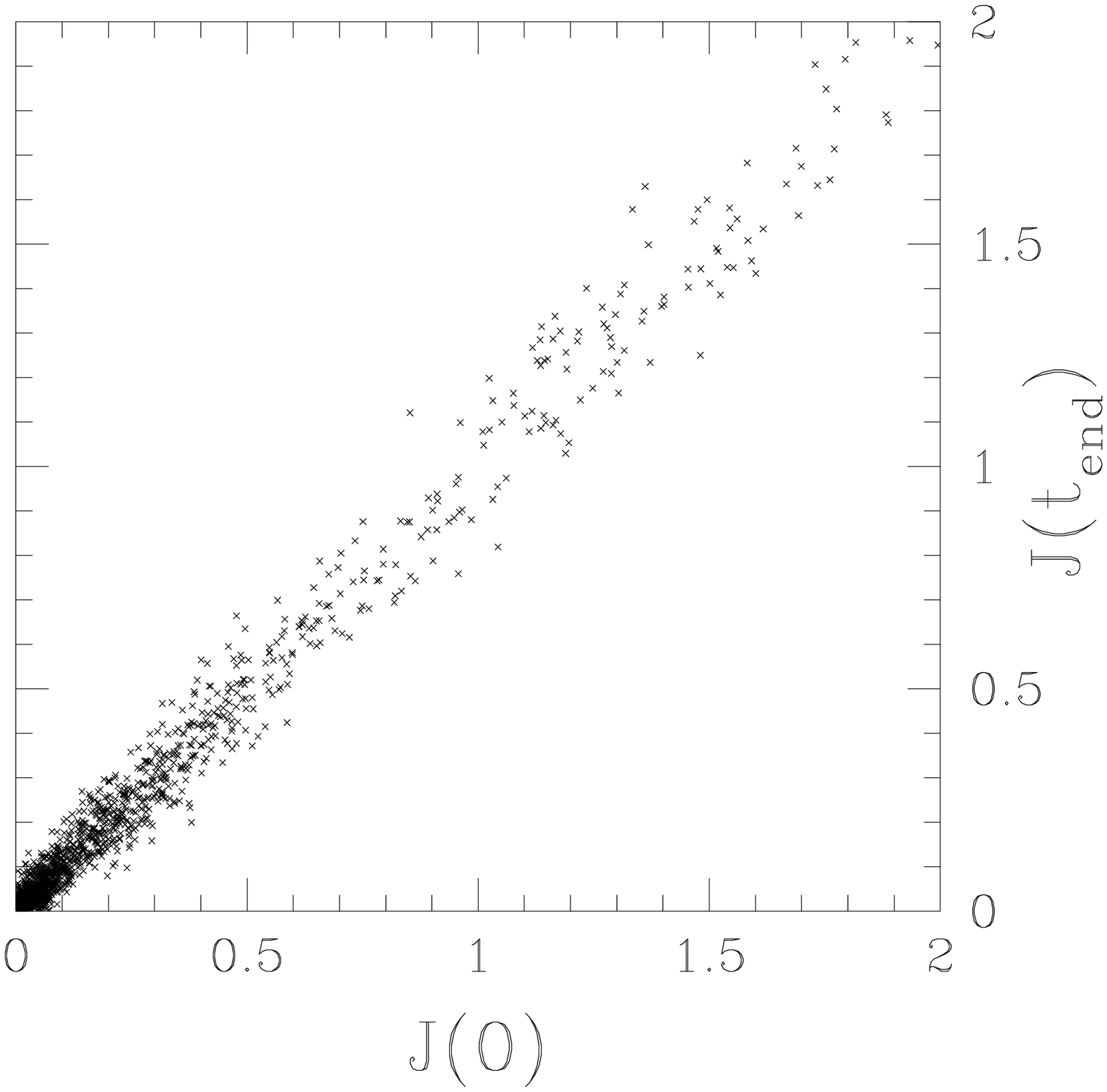}}
  \end{center}
  \caption{Single-particle angular momentum scatter (correlation
  between the initial and final values of $J$) for a typical clumpy
  simulation (left panel) and for its symmetrized (homogeneous)
  version (right panel). The symmetrization process in the initial
  state is performed by accepting the radius and the magnitude of the
  velocity of each particle, following the procedure for initial
  clumpy conditions, but by redistributing uniformly the angular
  variables.}
  \label{fig:jscatter}
\end{figure}

Earlier investigations~\cite{van82,lon91} compared ``clumpy" to
``homogeneous" initial conditions, showing that clumpy initial
conditions lead to end-states with projected density distributions
well fitted by the $R^{1/4}$ law. We argue that, if the degree of
symmetry in the initial conditions is excessive, little room is left
for the mechanism of violent relaxation; this is confirmed by the fact
that little or no mixing is observed in the single-particle angular
momentum for homogeneous simulations, as reported in
Fig.~\ref{fig:jscatter}. We have thus chosen to focus this paper on
clumpy initial conditions, for which an efficient mixing in phase
space is guaranteed. 

We performed several runs varying the number of clumps and the initial
virial ratio $u$ in the range from $0.05$ to $0.25$. In most cases the
clumps are cold, i.e. their kinetic energy is all associated with the
motion of their centers of mass. The simulations are run for several
dynamical times beyond the violent collapse phase, ensuring that the
system has settled into a quasi-equilibrium. The final configurations
are quasi-spherical, with shapes that resemble those of $E2-E3$
galaxies. The final equilibrium half-mass radius $r_M$ is basically
independent of the value of $u$. The central concentration achieved
correlates with $u$, as can be inferred from the conservation of
maximum density in phase space~\cite{lon91}.

\section{Fits and phase-space properties} \label{sec:fit}

In order to study the output of the simulations we compare the density
and the anisotropy profiles, $\rho (r)$ and $\alpha (r)$ (here the
local anisotropy is defined as $\alpha(r) =
2 - (\langle p^2_{\theta}\rangle + \langle p^2_{\phi} \rangle)/\langle
p^2_r \rangle$), of the end-products with the theoretical profiles of
the $f^{(\nu)}$ family of models. Smooth simulation profiles are
obtained by averaging over time in the last few dynamical times of the
simulation. For the fitting models, the parameter space explored is
that of an equally spaced grid in $(\nu,\Psi)$, with a subdivision of
$1/8$ in $\nu$, from $3/8$ to $1$, and of $0.2$ in $\Psi$, from $0.2$
to $14.0$; the mass and the half-mass radius of the models are fixed
by the scales set by the simulations. A minimum $\chi^2$ analysis is
performed as described elsewhere~\cite{ber03b}.

\begin{figure}
\includegraphics[height=.30\textheight]{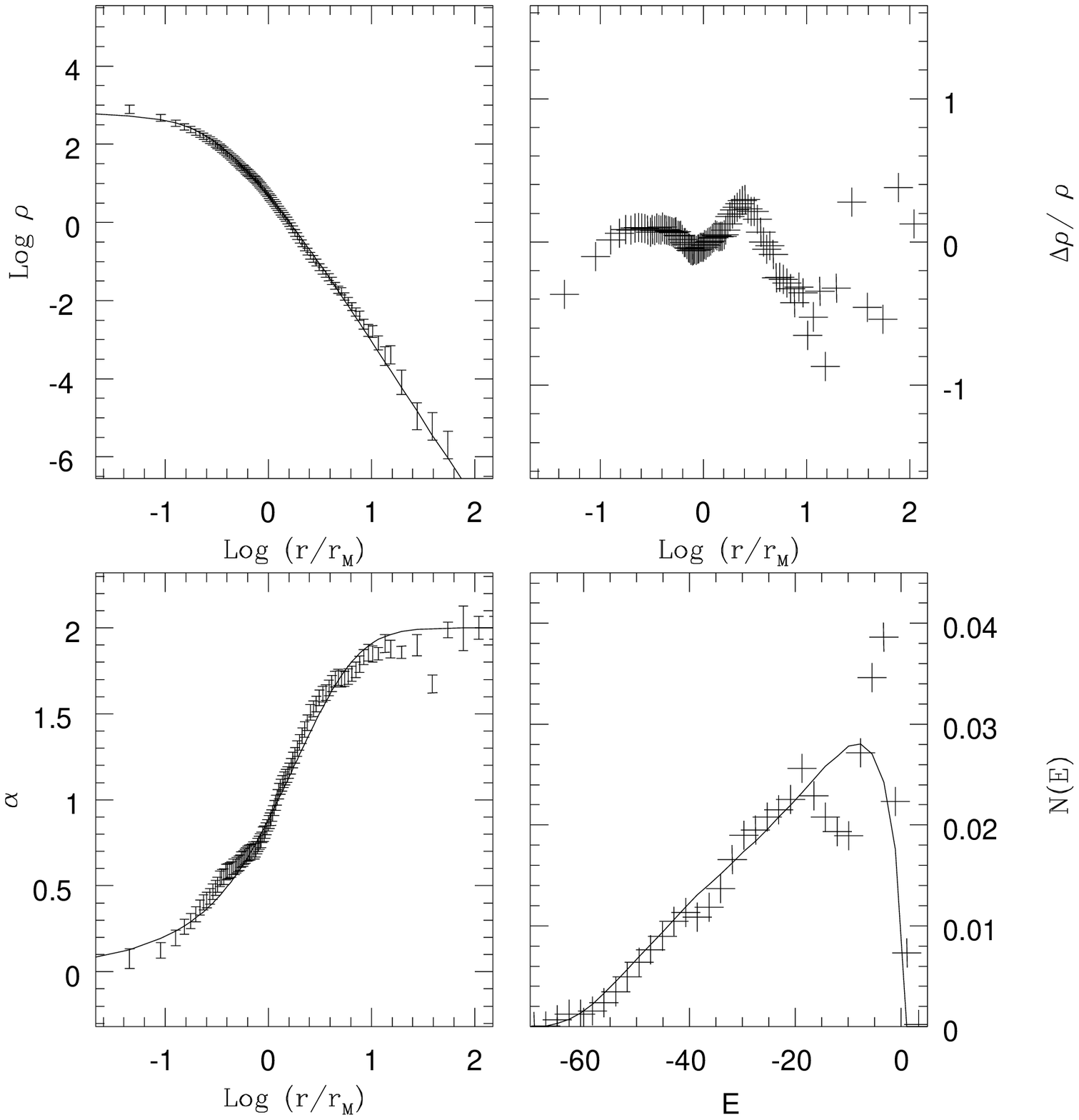}
\includegraphics[height=.28\textheight]{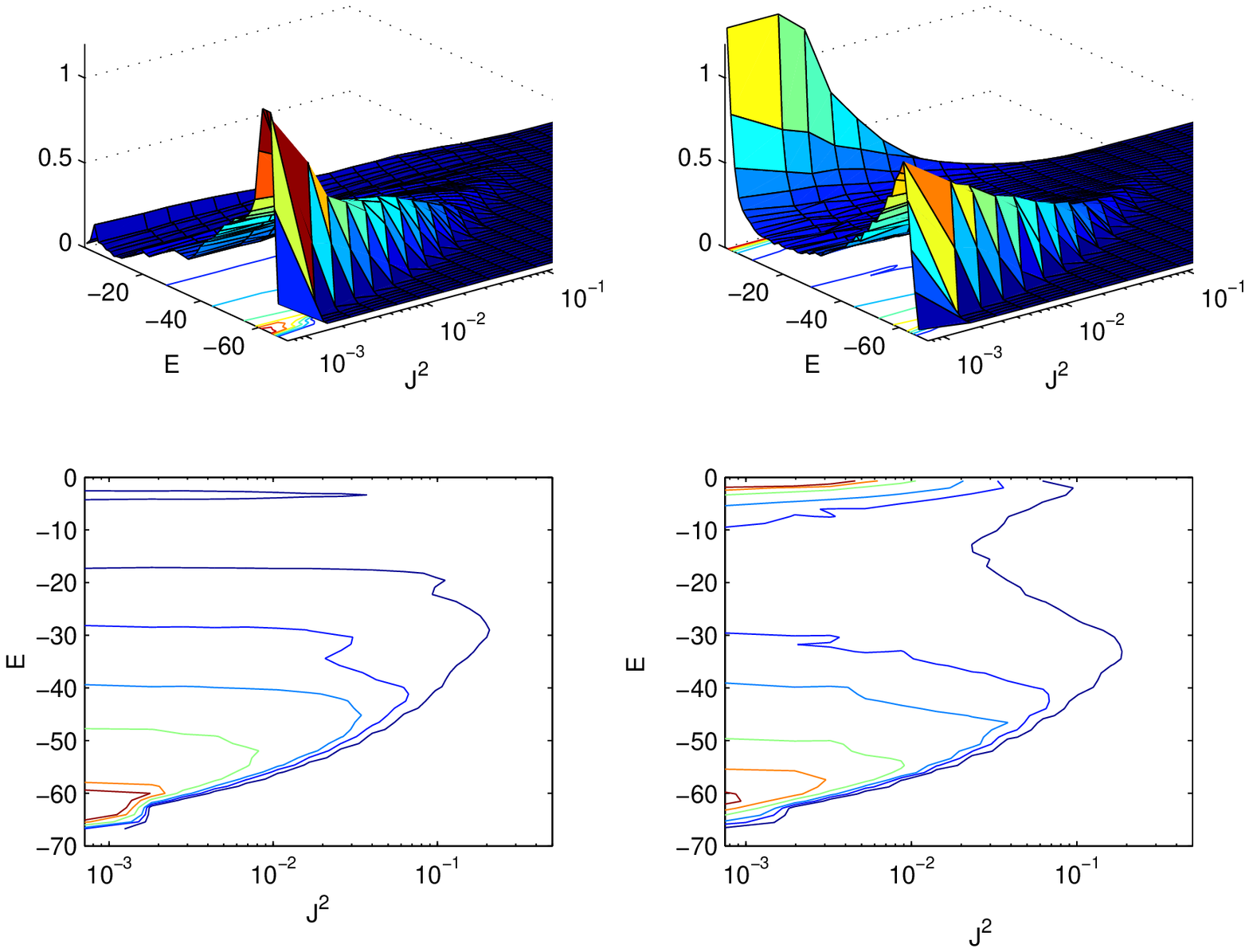}
\caption{Comparison between the results of one simulation with $u =
  0.23$, called $C2\_4$ (starting from 20 cold clumps), and the
  best-fit $f^{(\nu)}$ model ($\nu = 5/8$, $\Psi = 5.4$). {\it Left
    set of four panels}. Density as measured from the simulation
  (error bars) and model profile (top left).  Residuals from the
  best-fit profile (top right). Anisotropy profile of the simulation
  (error bars) and profile (bottom left).  Energy density distribution
  $N(E)$ (bottom right). {\it Right set of four panels}. Final phase
  space density $N(E,J^2)$ (left column), compared with that of the
  best fitting $f^{(\nu)}$ model (right column). The model curve for
  $N(E)$ and surface for $N(E,J^2)$ have been computed by a Monte
  Carlo sampling of phase space. \label{fig:simC2}}

\end{figure}

As illustrated in Fig.~\ref{fig:simC2}, the density of the simulations
is well represented by the best-fit $f^{(\nu)}$ profile over the
entire radial range. The fit is satisfactory not only in the outer
parts, where the density falls by {\it nine orders of magnitude} with
respect to the center, but also in the inner regions. Depending on the
initial virial parameter $u$, the end-products possess a density
profile which, projected along the line-of-sight, may exhibit an
$R^{1/n}$ behavior with different values of $n$ and yet it is well
fitted by the $\f$ models. In turn, this may be interpreted in the
framework of the proposed weak homology of elliptical
galaxies~\cite{ber02}.

To some extent, the final anisotropy profiles for clumpy initial
conditions are found to be sensitive to the detailed choice of
initialization, in particular to the number and the size of the
clumps. The agreement between end-products of the collapse and models
(see Fig.~\ref{fig:simC2}) seems to be best for simulations with $10$
to $20$ clumps, for a clump size such that the sum of their volumes
fills the sphere where their centers of mass lie. We should recall
that both the low and the high number of clumps limits are those of an
initial homogeneous condition, unfavorable to violent relaxation.

At the level of phase space, we have performed two types of
comparison, one involving the energy density distribution $N(E)$ and
the other based on $N(E,J^2)$. The chosen normalization factors are
such that: $M = \int N(E) dE = \int N(E,J^2) dE dJ^2.$ In
Fig.~\ref{fig:simC2} we plot the final energy density distribution for
a simulation run called $C2\_4$ with respect to the predictions of the
best-fit model identified from the study of the density and pressure
anisotropy distributions. The agreement is good, especially for
strongly bound particles. At the deeper level of $N(E,J^2)$,
simulations and models also agree quite well, as illustrated in the
right set of four panels of Fig.~\ref{fig:simC2}. For the case shown,
the distribution contour lines basically match in the range from
$E_{min}$ to $E \approx - 30$ (in the relevant units); however, the
theoretical model shows a peak located near the origin, not observed
in the simulations.

\end{document}